\documentclass[aps, prd, amsmath, floats, floatfix, twocolumn, nofootinbib, superscriptaddress]{revtex4-1}
\usepackage[pdftex]{graphicx}
\usepackage{color}
\usepackage{latexsym}
\usepackage[colorlinks]{hyperref}
\usepackage{amsthm,amsmath,amssymb}
\usepackage{mathrsfs}
\usepackage{makecell}
\usepackage{natbib}
\usepackage[normalem]{ulem}

\newcommand{\be}{\begin{eqnarray}}
\newcommand{\ee}{\end{eqnarray}}

\newcommand{\apjl}{Astrophys.\ J.\ }

\newcommand{\aap}{Astron.\ Astrophys.\ }
\newcommand{\araa}{Annu.\ Rev.\ Astron. Astrophys.\ }

\newcommand{\mnras}{Mon.\ Not.\ R.\ Astron.\  Soc.\ }

\newcommand{\ssr}{Space Sci.\ Rev.\ }

\begin{document}

\title{Assessing the Power-Law Emissivity Assumption in X-ray Reflection Spectroscopy:\\A Simulation-Based Evaluation of Different Coronal Geometries}

\author{Songcheng~Li}
\affiliation{Center for Astronomy and Astrophysics, Department of Physics, Fudan University, Shanghai 200438, China}

\author{Abdurakhmon~Nosirov}
\affiliation{Center for Astronomy and Astrophysics, Department of Physics, Fudan University, Shanghai 200438, China}
\affiliation{Institut f\"ur Astronomie und Astrophysik, Eberhard-Karls Universit\"at T\"ubingen, D-72076 T\"ubingen, Germany}

\author{Cosimo~Bambi}
\email[Corresponding author: ]{bambi@fudan.edu.cn}
\affiliation{Center for Astronomy and Astrophysics, Department of Physics, Fudan University, Shanghai 200438, China}
\affiliation{School of Natural Sciences and Humanities, New Uzbekistan University, Tashkent 100000, Uzbekistan}

\author{Honghui~Liu}
\affiliation{Institut f\"ur Astronomie und Astrophysik, Eberhard-Karls Universit\"at T\"ubingen, D-72076 T\"ubingen, Germany}

\author{Zuobin~Zhang}
\affiliation{Astrophysics, Department of Physics, University of Oxford, Oxford OX1 3RH, UK}

\author{Shafqat~Riaz}
\affiliation{Theoretical Astrophysics, Eberhard-Karls Universität Tübingen, D-72076 Tübingen, Germany}

\begin{abstract}
The emissivity profile assumed in X-ray reflection spectroscopy significantly impacts black hole spin measurements. Using simulated NuSTAR spectra generated for lamppost and disk-like coronae with the {\tt relxill} model suite, we evaluate systematic biases introduced when fitting with power-law or broken power-law emissivity profiles. We find that a simple power-law can accurately recover spins for low-height lamppost coronae with long exposures and low inclination angle, while broken power-laws introduce degeneracies when the simple power-law already performs adequately. However, for extended or high-height coronae, especially at high inclinations, both models produce large systematic biases unresolved by longer exposure times. Our results demonstrate that power-law approximations are reliable for compact coronae, highlighting the need for geometry-specific models in complex cases.
\end{abstract}

\maketitle


\section{Introduction}

X-ray reflection spectroscopy is a powerful tool for measuring the spins of black holes in X-ray binaries and active galactic nuclei and for studying the innermost structure of accretion disks~\cite{Miller_2007ARA&A..45..441M, Miller_2009ApJ...697..900M, Reynolds_2014SSRv..183..277R,Bambi_2021SSRv..217...65B}. The reflection spectrum is produced when the inner accretion disk is illuminated by hard X-rays from a ``corona''~\cite{Galeev_1979ApJ...229..318G, Haardt_Maraschi_1993ApJ...413..507H}, which is some hotter (electron temperature $kT_{\rm e} \sim 100$~keV), usually optically-thin, plasma near the black hole and the inner part of the accretion disk. 

The emissivity profile, normally indicated as $\epsilon(r)$, describes the irradiating intensity distribution across the disk and is an important ingredient in the analysis of the reflection features of a source~\cite{Wilkins_2011MNRAS.414.1269W}. The emissivity profile is determined by the properties of the corona (geometry, velocity, emissivity, etc.) and determines the observable reflection spectral features. If we knew the properties of the corona, the intensity profile could be computed~\cite{2024arXiv240812262B}. For example, in the lamppost model, the corona is a point-like source along the black hole spin axis. The emissivity profile of a lamppost corona can be calculated as a function of the radial coordinate $r$ and the corona height $h$~\cite{Dauser_2013MNRAS.430.1694D}. As demonstrated in Ref.~\cite{Dauser_2014MNRAS.444L.100D}, if the system has a lamppost corona and we employ a reflection model with a lamppost emissivity profile, we infer significantly more precise and accurate constraints on the black hole parameters such as the spin, while these constraints become inaccurate if the assumed coronal geometry is incorrect. 

Unfortunately, the coronal geometry is normally unknown. It may exhibit a diverse range of morphologies. In Ref.~\cite{Gonzalez_2017MNRAS.472.1932G}, the authors show that the geometry of the corona cannot be completely distinguished merely by the emissivity profile and additional auxiliary indicators are also required. This is because there is not a one-to-one correspondence between the corona and the emissivity profile: the same emissivity profile can be produced by coronae with different geometries.

Since the coronal geometry is normally unknown, in data analysis it is common to model the emissivity profile with a power-law ($\epsilon \propto r^{-q}$, where $q$ is the emissivity index) or a broken power-law ($\epsilon \propto r^{-q_{\rm in}}$ for $r < R_{\rm br}$ and $\epsilon \propto r^{-q_{\rm out}}$ for $r > R_{\rm br}$, where $q_{\rm in}$, $q_{\rm out}$, and $R_{\rm br}$ are, respectively, the inner emissivity index, the outer emissivity index, and the breaking radius) and infer the values of $q$, $q_{\rm in}$, $q_{\rm out}$, and $R_{\rm br}$ directly from the fit. However, this is clearly an approximation and can introduce undesired systematic biases in the estimates of the properties of a source. With such an approximation, we also ignore a number of relativistic effects (e.g., the photon redshift between the corona and the accretion disk)~\cite{2024arXiv240812262B}, which may not be completely negligible in the presence of high-quality data~\cite{2025MNRAS.536.2594L}.

The power-law emissivity model remains extensively adopted in observational studies of both X-ray binaries and active galactic nuclei~\cite{Bambi_2021SSRv..217...65B}. This practice stems from its simplicity and lower computational cost but introduces critical and unquantified risks. First, the empirically emissivity indices inferred from a fit may be incorrectly correlated with the geometrical parameters; see the discussion in Ref.~\cite{Szanecki_2020A&A...641A..89S}. Second, we may have systematic biases in the estimates of the black hole parameters, such as the black hole spin parameter $a_{*}$, when the emissivity profile of the true coronal geometry deviates significantly from the power-law form. In a broken power-law profile, steep emissivity indices are commonly recovered in the innermost accretion disk (see, e.g., Refs.~\cite{El-Batal_2016ApJ...826L..12E,Draghis_2023ApJ...946...19D}). In the inner part of the disk, a compact corona enhances the inner-disk illumination via light bending, producing steep emissivity slopes, while an extended corona generates flatter profiles~\cite{Wilkins_and_Fabian_2012MNRAS.424.1284W}. However, inaccurate modeling of the emissivity profile does not necessarily lead to inaccurate spin measurements despite errors in some other parameters. We need a systematic assessment of how the power-law approximation performs across diverse coronal geometries and its impact on parameter estimations. In the case of high-quality data, the quality of the fit should indicate whether the emissivity profile model is suitable for describing our astrophysical system. Fits employing inappropriate emissivity profiles may yield incorrect measurements of certain model parameters (see, e.g., the discussion in Refs.~\cite{Zhang_2019ApJ...884..147Z,Liu_2019PhRvD..99l3007L}).

Previous studies have already investigated the influence of various coronal geometries and emissivity profiles on spin measurements; see, for instance, Ref.~\cite{Zhang_2024MNRAS.532.3786Z}. However, none of them has quantified the errors induced by forcing a power-law (or broken power-law) profile onto spectra generated from physically motivated coronae. This gap may obscure the uncertainty margins in key black hole parameters such as the spin derived from observation studies. To address this issue, in this work we conduct a controlled simulation experiment using synthetic spectra generated with XSPEC. Our goals are threefold: $i)$ quantifying spin biases when fitting power-law emissivity models to spectra produced by distinct coronal geometries, $ii)$ linking the sensitivity of these biases to the coronal properties (height, size), black hole properties (spin, inclination angles), and observational conditions (exposure time), and $iii)$ providing guidelines for identifying scenarios where power-law fits remain reliable versus those requiring self-consistent geometric modeling.

Coronae can exhibit various geometric configurations, including lamppost, sandwich, spherical, and toroidal structures; see, for instance, Refs.~\cite{2024arXiv240812262B,Bambi_2017bhlt.book.....B} and references therein. In this paper, we select the two simplest and most representative static configurations: lamppost and disk-like. 
Disk-like coronae are simple extension of the lamppost model to give the corona a finite size in the transverse direction~\cite{Miniutti_2003MNRAS.344L..22M, Suebsuwong_2006A&A...453..773S, Wilkins_and_Fabian_2012MNRAS.424.1284W}.
These models effectively illustrate how both the vertical distance from the black hole and the radial extent of the corona influence the emissivity profile.

We generate high-quality NuSTAR spectra using {\tt relxill\_nk} (specifically, {\tt relxilllp\_nk} and {\tt relxilldisk\_nk}~\cite{Bambi_2017ApJ...842...76B, relxillnk_2019ApJ...878...91A,2020ApJ...899...80A})\footnote{Even if in our study we assume the Kerr spacetime, we use {\tt relxill\_nk} instead of {\tt relxill}~\cite{relxill_2014ApJ...782...76G} because {\tt relxill} does not have a model for a disk-like corona and a model for a twice-broken power-law emissivity profile.}. Employing input spectra generated by specific coronal geometries, we fit the simulated spectra with the standard power-law or broken power-law emissivity models. By comparing input and output parameters, we rigorously evaluate the efficacy and limitations of the power-law assumption. This approach circumvents the challenge of requiring a prior knowledge of coronal geometry in actual observations, thereby providing robust guidance for future observational and theoretical studies of X-ray reflection spectroscopy.

The content of the manuscript is as follows. In Section~\protect\ref{data simulation}, we present our simulation setup and fitting procedures. In Section~\ref{results}, we present the parameter estimation biases across the lamppost and disk-like geometries. We discuss our results in Section~\ref{discussion} and we report our conclusions in Section~\ref{conclusion}.

\section{Simulations}
\label{data simulation}

Using XSPEC~\cite{xspec_1996ASPC..101...17A} version 12.13.0c, we simulate NuSTAR~\cite{NuSTAR_Harrison_2013ApJ...770..103H} spectra of bright black hole X-ray binaries at the end of the hard state, when the luminosity of the source is high and we can expect that the inner edge of the disk is at the radius of the innermost stable circular orbit (ISCO). We assume that the flux of the source is $2\times10^{-8}$~erg~s$^{-1}$~cm$^{-2}$ in the energy band 2-10~keV. We set the exposure time 30~ks and 300~ks for the focal plane module A (FPMA). Each spectrum is generated via the XSPEC {\tt fakeit} command, employing standard response files from the NuSTAR official website\footnote{\href{https://nustar.caltech.edu/page/response-files}{https://nustar.caltech.edu/page/response-files}}: the response matrix file (RMF) {\tt nustar.rmf}, the ancillary response file (ARF) {\tt point\_60arcsecRad\_1arcminOA.arf}, and the background spectrum {\tt bgd\_60arcsec.pha}.

We simulate the spectra using {\tt relxill\_nk}~\cite{Bambi_2017ApJ...842...76B, relxillnk_2019ApJ...878...91A,2020ApJ...899...80A}, which is an extension of the {\tt relxill} package~\cite{relxill_2014ApJ...782...76G} and is public on GitHub\footnote{\href{https://github.com/ABHModels/relxill_nk}{https://github.com/ABHModels/relxill\_nk}}. In {\tt relxill\_nk}, we have the lamppost model {\tt relxilllp\_nk}~\cite{relxillnk_2019ApJ...878...91A} and the model for a disk-like corona {\tt relxilldisk\_nk}~\cite{relxilldisknk_2022ApJ...925...51R}.

We start with the lamppost model. The XSPEC model for our simulations is {\tt tbabs*relxilllp\_nk}, where {\tt tbabs} describes the Galactic absorption with the abundance table from Ref.~\cite{angr_1989GeCoA..53..197A}. {\tt tbabs} has only one parameter, the hydrogen column density: we set it to $N_{\rm H} = 0.6$ (in units of $10^{22}$~atoms~cm$^{-2}$). In {\tt relxilllp\_nk}, we set the following values to the parameters of the model: photon index $\Gamma = 1.7$, high-energy cutoff $E_{\rm cut} = 300$~keV, ionization parameter of the disk $\log\xi = 3.1$ ($\xi$ in erg~cm~s$^{-1}$), iron abundance $A_{\rm Fe} = 1$ (Solar iron abundance), and reflection fraction $R_{\rm F} = 1$ to make the coronal intensity illuminating the disk equal to that escaping to infinity\footnote{We note that in the presence of a specific coronal geometry, the reflection fraction is not a free parameter and can be properly estimated by ray-tracing calculation~\cite{Dauser_2014MNRAS.444L.100D}. However, it is enough that the coronal emission is not isotropic, the corona is not static, etc. that the value of this reflection fraction can appreciably changes.} (see Ref.~\cite{Dauser_2016A&A...590A..76D} for the definition of $R_{\rm F}$). We assume that the spacetime geometry is described by the Kerr solution and we consider two possible values for the inclination angle of the disk, namely the angle between the black hole spin axis and the line of sight of the observer: $i = 20^\circ$ and $70^\circ$. In the lamppost model, the corona has only one parameter, the height $h$: we consider two values, $h = 4$~$r_g$ and $h = 10$~$r_g$, where $r_g = M$ is the gravitational radius of the black hole. For the spin, we consider 11~values: $a_* = 0$, 0.1, 0.2, 0.3, 0.4, 0.5, 0.6, 0.7, 0.8, 0.9, and 0.998. 

For the disk-like corona setup, the XSPEC model is {\tt tbabs*relxilldisk\_nk}. Here, the corona is an annulus with inner radius $R_{\rm disk,in}$, outer radius $R_{\rm disk,out}$, and is parallel to the equatorial plane at the height $h$. We fix $R_{\rm disk,in} = 1$~$r_g$ (which is the minimum value allowed by the model). In order to study how the compactness of a corona affects the spin measurements, we consider three possible values for the outer radius: $R_{\rm disk,out} = 2$, 5, and 20~$r_g$. For the height of the corona, we consider two possible values: $h = 4$~$r_g$ and $h = 10$~$r_g$. All other parameters match the lamppost configuration. It should be noted that in {\tt relxilldisk\_nk}, the coronal geometry is only used to compute the corresponding emissivity on the disk~\cite{relxilldisknk_2022ApJ...925...51R}. In this study, we do not consider the effect whereby an extended corona (for example, disk-like corona with large $R_{\rm disk,out}$) may scatter reflection photons along their path from the disk to the observer~\cite{2017ApJ...836..119S}, since including this reflection Comptonization effect is unlikely to produce significant differences in spectral analysis, as indicated in Refs.~\cite{2015MNRAS.449..129W,Li_2024PhRvD.110d3021L}.

Finally, we grouped all the spectra with the optimal binning algorithm in \cite{Kaastra_2016A&A...587A.151K} by using the {\tt ftgrouppha} task.

\section{Data Analysis and Results}
\label{results}

We utilize XSPEC version 12.13.0c for spectral analysis, performing fits across the 3-79~keV energy band. We fit all simulated spectra with the XSPEC model {\tt tbabs*relxill\_nk} assuming the Kerr spacetime. For every case, first we fit the data with the command {\tt fit}, then we run {\tt error}, {\tt steppar}, and {\tt error} again in order find the actual best-fit and parameter uncertainties. We fix the hydrogen column density to $N_{\rm H} = 0.6$. For the emissivity profile, first we employ a simple power-law, so we have only a free parameter, the photon index $q$. After the power-law profile, we try the broken power-law, where there are three parameters: the inner emissivity index $q_{\rm in}$, the outer emissivity index $q_{\rm out}$, and the breaking radius $R_{\rm br}$. We also examine the twice-broken power-law profile with five parameters: the emissivity index of the inner part of the disk $q_{1}$, the emissivity index of the central part of the disk $q_{2}$, the emissivity index of the outer part of the disk $q_{3}$, and two breaking radii $R_{\rm br1}$ and $R_{\rm br2}$. We fix the inner radius of the disk $R_{\rm in}$ to the ISCO, the outer radius of the disk $R_{\rm out}$ to $400$~$r_g$, and the redshift to $z = 0$. All other parameters remain free in the fits. We use the $\chi^2$ statistics to find the best-fit model.

\subsection{Lamppost Corona}
\label{results_lamppost}

\begin{figure*}[t]
	\centering
	\includegraphics[width=\linewidth]{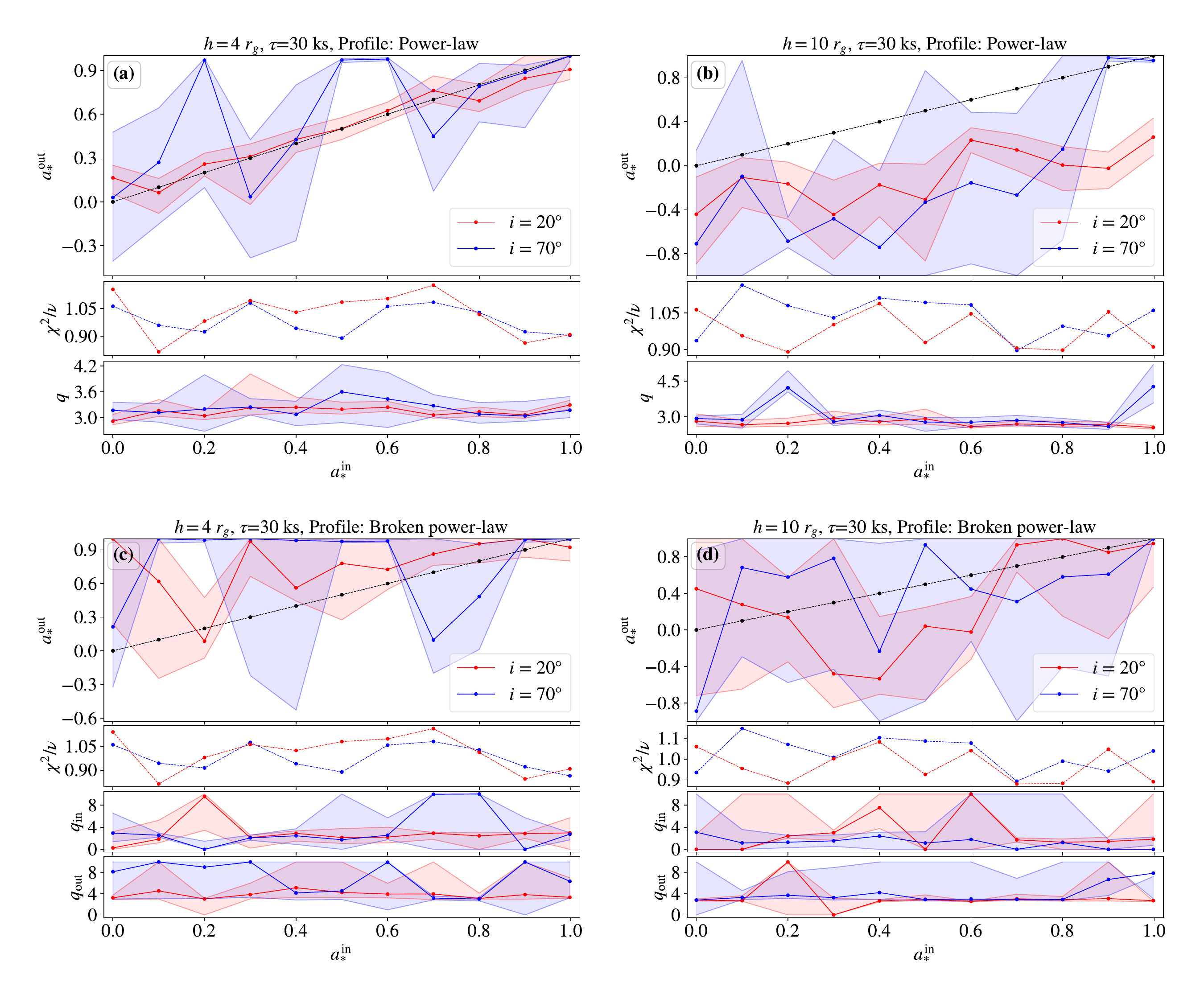}
	\vspace{-1.0cm}
	\caption{Results for the lamppost corona simulations with corona height $h=4$~$r_{g}$ (left panels) and $h=10$~$r_{g}$ (right panels) when the exposure time is $\tau=30$~ks. In the top panels, we fit the data with a power-law emissivity profile. In the bottom panels, we fit the data with a broken power-law emissivity profile. In every panel, we show: input spin $a_*^{\rm in}$ vs output spin $a_*^{\rm out}$ (top quadrant), reduced $\chi^{2}$ of the best-fit model (central quadrant), and input spin vs best-fit emissivity index/indices (bottom quadrants). The dots indicate the best-fit values and the shaded regions indicate the 90\% confidence intervals. Red is for the simulations with input inclination angle $i=20^\circ$ and blue is for the simulations with $i=70^\circ$. In the quadrants input spin vs output spin, we report in black the reference line $a_*^{\rm in} = a_*^{\rm out}$.}
	\label{nustar_30ks_lampost}
\end{figure*}

For the simulations with an exposure time of 30~ks, the results are summarized in Fig.~\ref{nustar_30ks_lampost}. The input coronal height is $h = 4$~$r_g$ (left panels) and $h = 10$~$r_g$ (right panels). In the fits, we employ a simple power-law emissivity profile in the top panels and a broken power-law emissivity profile in the bottom panels. In general, the assumption of a simple power-law provides more accurate spin measurements than the choice of a broken power-law, at least for $h = 4$~$r_g$ (while for $h = 10$~$r_g$ we obtain inaccurate results in both cases). We thus repeat the fits with a broken power-law emissivity profile and imposing $q_{\rm out} = 3$, which corresponds to the Newtonian limit at large radii for a lamppost corona. This choice does not lead to any significant improvement of spin measurements and therefore we do not report the results here.

\begin{figure*}[t]
	\centering
	\includegraphics[width=0.9\linewidth]{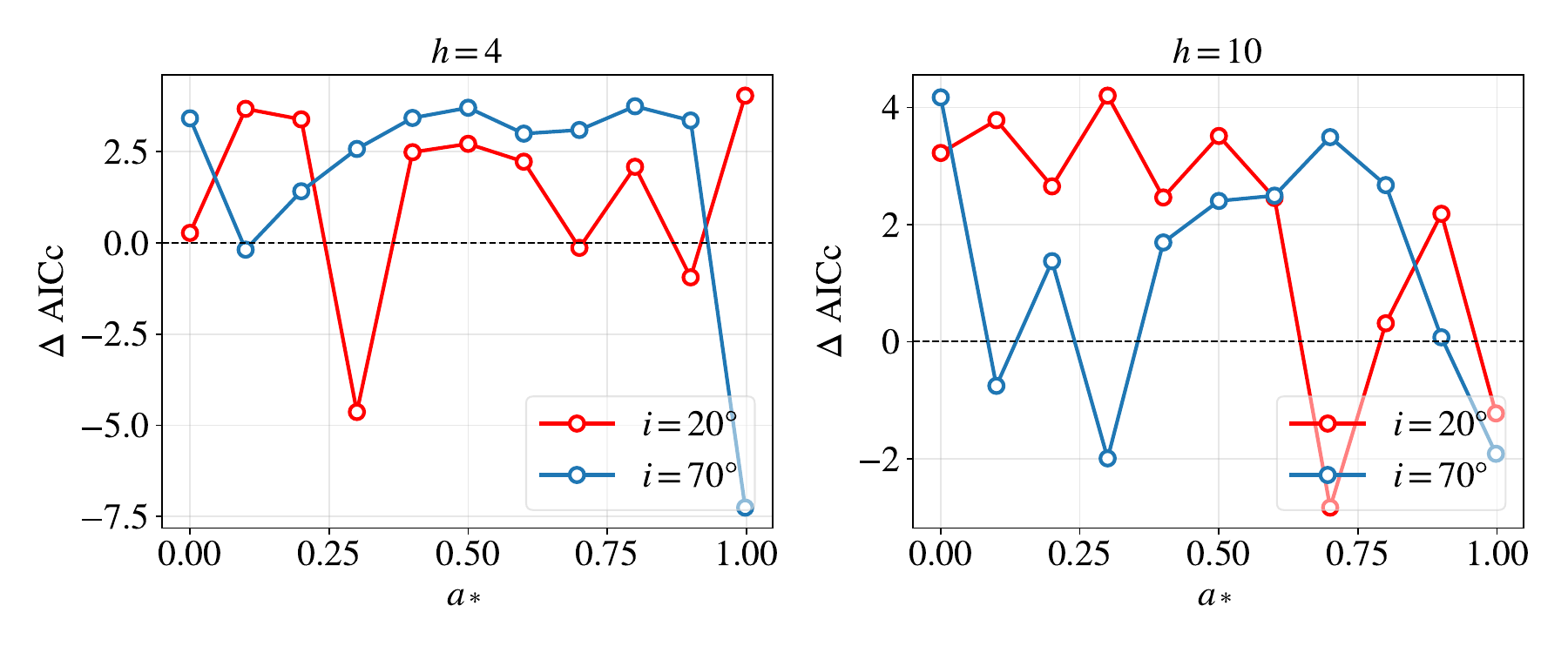}
	\vspace{-0.6cm}
	\caption{Comparison of the quality of the fits with power-law emissivity profile and broken power-law emissivity profile in Fig.~\ref{nustar_30ks_lampost} with the Akaike information criterion corrected for small sample sizes. $\Delta$AICc = AICc (broken power-law) -- AICc (power-law).}
	\label{f-AICc}
\end{figure*}

To determine which model, between the simple power-law emissivity profile and the broken power-law emissivity profile, fits the data best, we apply the Akaike information criterion corrected for small sample sizes (AICc)~\cite{1974ITAC...19..716A}, which is a most robust method than the comparison of the minimum of $\chi^2$. We use the following formula to calculate AICc
\be
{\rm AICc} = \chi^2_{\rm min} + 2N_p + \frac{2 N_p \left( N_p + 1\right)}{\left( N_b - N_p - 1 \right)} \, , 
\ee
where $\chi^2_{\rm min}$ is the minimum of $\chi^2$, $N_p$ is the number of free parameters, and $N_b$ is the number of bins. 
AICc assesses only relative predictive performance, not physical correctness. The model with the lowest AICc is favored by the data. If the difference of the AICc value between two models is in the range 0-5, the quality of the fits is similar; if the difference is in the range 5-10, the model with higher AICc is less favored by the data; if the difference exceeds 10, the model with higher AICc can be ruled out. As we can see in Fig.~\ref{f-AICc}, the fits with a simple power-law emissivity profile have normally a lower AICc (especially for $h = 4$~$r_g$), but the difference of the AICc value is always less than 5 (with the exception of one case).

\begin{figure*}[t]
	\centering
	\includegraphics[width=\linewidth]{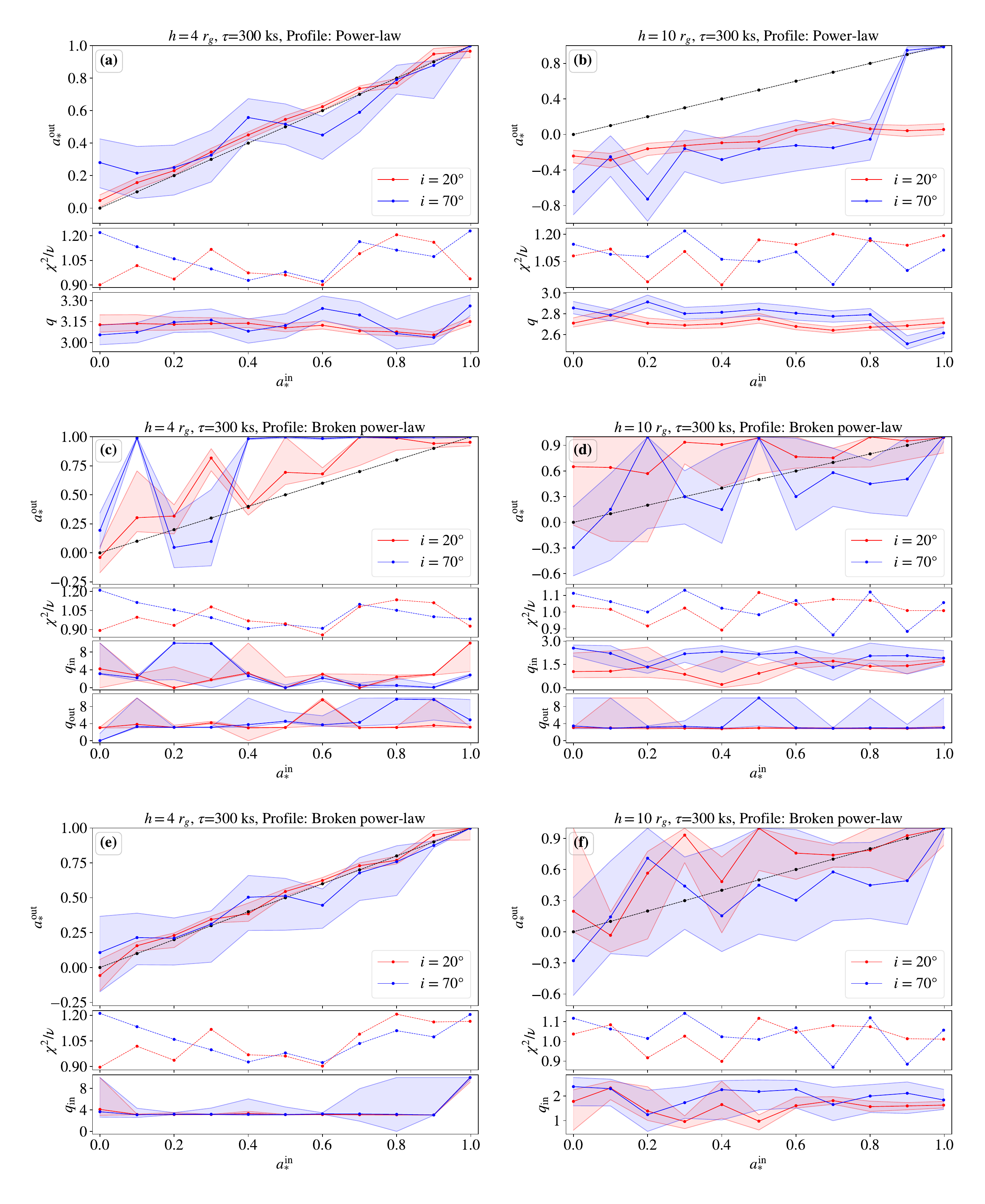}
	\vspace{-1.0cm}
	\caption{As in Fig.~\ref{nustar_30ks_lampost} for simulations with exposure time $\tau=300$~ks. In the top panels, we fit the data with a power-law emissivity profile. In the central panels, we fit the data with a broken power-law emissivity profile. In the bottom panels, we fit the data with a broken power-law emissivity profile and we impose $q_{\rm out} = 3$.}
	\label{nustar_300ks_lampost}
\end{figure*}

The results of the simulations with exposure time 300~ks are summarized in Fig.~\ref{nustar_300ks_lampost}. As in Fig.~\ref{nustar_30ks_lampost}, the input coronal height is $h = 4$~$r_g$ in the left panels and $h = 10$~$r_g$ in the right panels. We fit the simulated data with a simple power-law emissivity profile in the top panels, a broken power-law emissivity profile in the central panels, and a broken power-law emissivity profile with $q_{\rm out} = 3$ in the bottom panels. We find that, for $h = 4$~$r_{g}$, using a broken power-law emissivity profile with fixed $q_{\rm out} = 3$ yields more accurate spin measurements (closer to the input values with narrower confidence intervals) compared to the fits with $q_{\rm out}$ left free. This improvement may result from reducing parameter degeneracy in the model fitting process.

Comparison with the above results of lamppost coronae reveals that for high corona height cases ($h=10$~$r_{g}$), the confidence intervals of spin measurements broaden compared to the cases with lower heights ($h=4$~$r_{g}$), accompanied by a slight systematic underestimation of spin values -- for example, compare the results in the top left panel and in the top right panel of Fig.~\ref{nustar_30ks_lampost}. Regarding exposure time, longer exposures lead to narrower confidence intervals and improved precision across all input spins for low-height coronae, with particularly enhanced measurements for high-spin systems -- for example, compare the results in the top left panel in Fig.~\ref{nustar_30ks_lampost} and those in the top left panel in Fig.~\ref{nustar_300ks_lampost}. The cases with lower inclination ($i=20^{\circ}$) angles generally yield more accurate spin measurements. In general, lamppost corona systems characterized by low coronal height, low inclination, high spin, and long exposure times produce more accurate constrained spin measurements with the power-law emissivity profile model.

\begin{figure}[t]
	\centering
	\includegraphics[width=\linewidth]{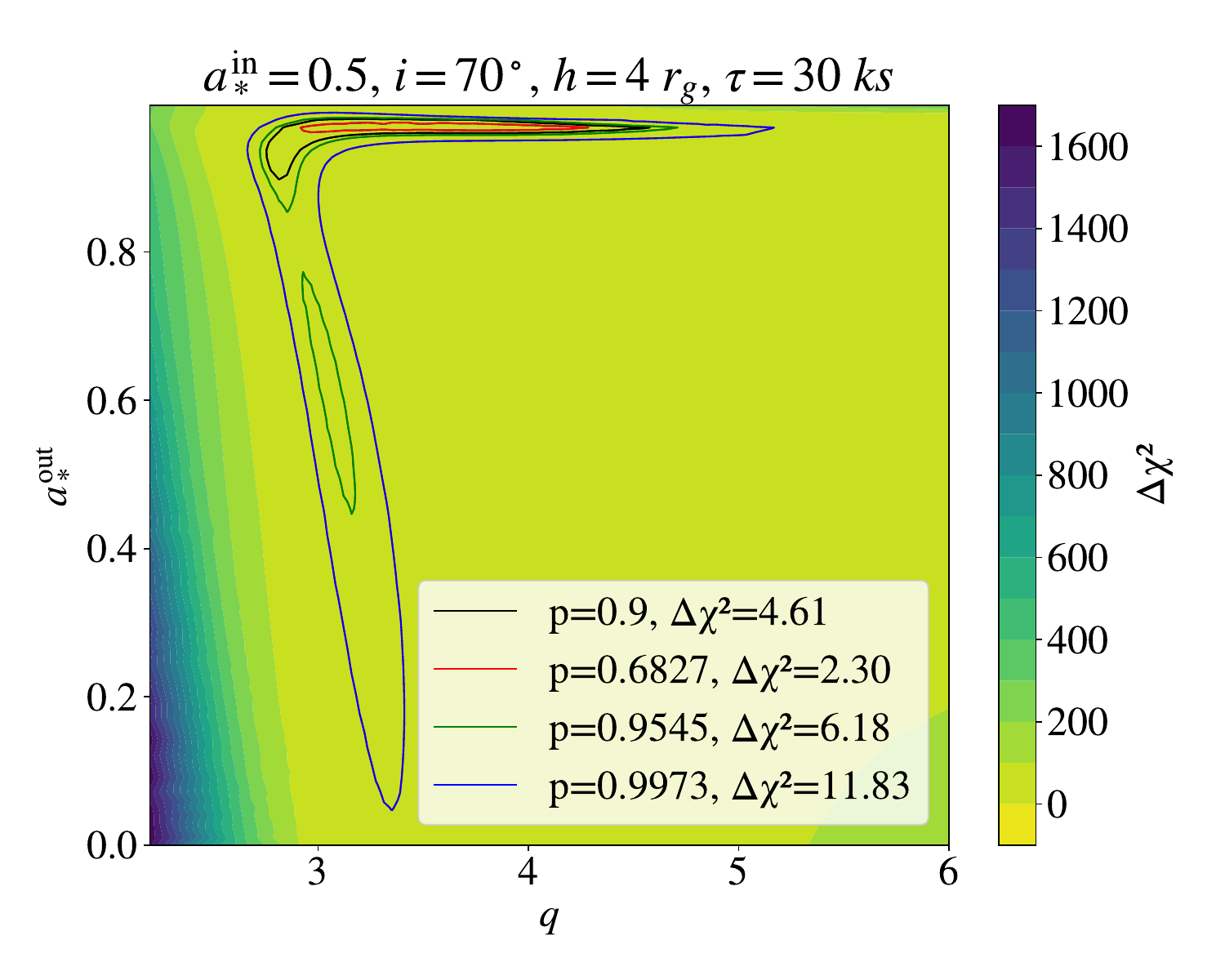}
	\vspace{-0.8cm}
	\caption{Contour plot of $\Delta\chi^2$ for spin ($a_*$) and emissivity index derived from {\tt steppar} calculation in XSPEC, for the case with input parameters $a_*^{\rm in} = 0.5$, $i = 70^\circ$, $h = 4~r_g$, and exposure time $\tau = 30$ ks, fitted with a simple power-law emissivity profile. The contours correspond to confidence levels of 1$\sigma$, 90\%, 2$\sigma$, and 3$\sigma$ for two parameters. The color scale represents $\Delta\chi^2$ values, with darker shades indicating higher $\Delta\chi^2$.
    }
	\label{contour_a_500}
\end{figure}

We also note that in some cases we have high spin measurements for some slow-rotating black holes with high inclination $i=70^{\circ}$. A more detailed study of those cases shows that their $\chi^2$ has the absolute minimum at $a_* = 0.998$ (which is the maximum value of the black hole spin allowed by the model), but there is also a local minimum with a slightly higher $\chi^2$ around the actual value of the spin parameter of the black hole of the simulation. For example, Fig.~\ref{contour_a_500} corresponds to the case shown in the top left panel of Fig.~\ref{nustar_30ks_lampost} with input spin $a_*^{\rm in} = 0.5$. Note that the input parameters are normally recovered well with the possible exception of the black hole spin parameter, so parameter degeneracy may only be between the spin parameter and the parameters related to the emissivity profile. The XSPEC {\tt error} command calculated a 90\% confidence interval around a high-spin absolute minimum. However, a local minimum also exists near the input spin value at the 2$\sigma$ (95.45\%) confidence level. Furthermore, at the 3$\sigma$ (99.73\%) level, the spin measurement spans a wide range and indicate a very broad confidence interval. This behavior is similar to the nearby case with $a_*^{\rm in} = 0.4$. 

\begin{figure*}[t]
	\centering
	\includegraphics[width=\linewidth]{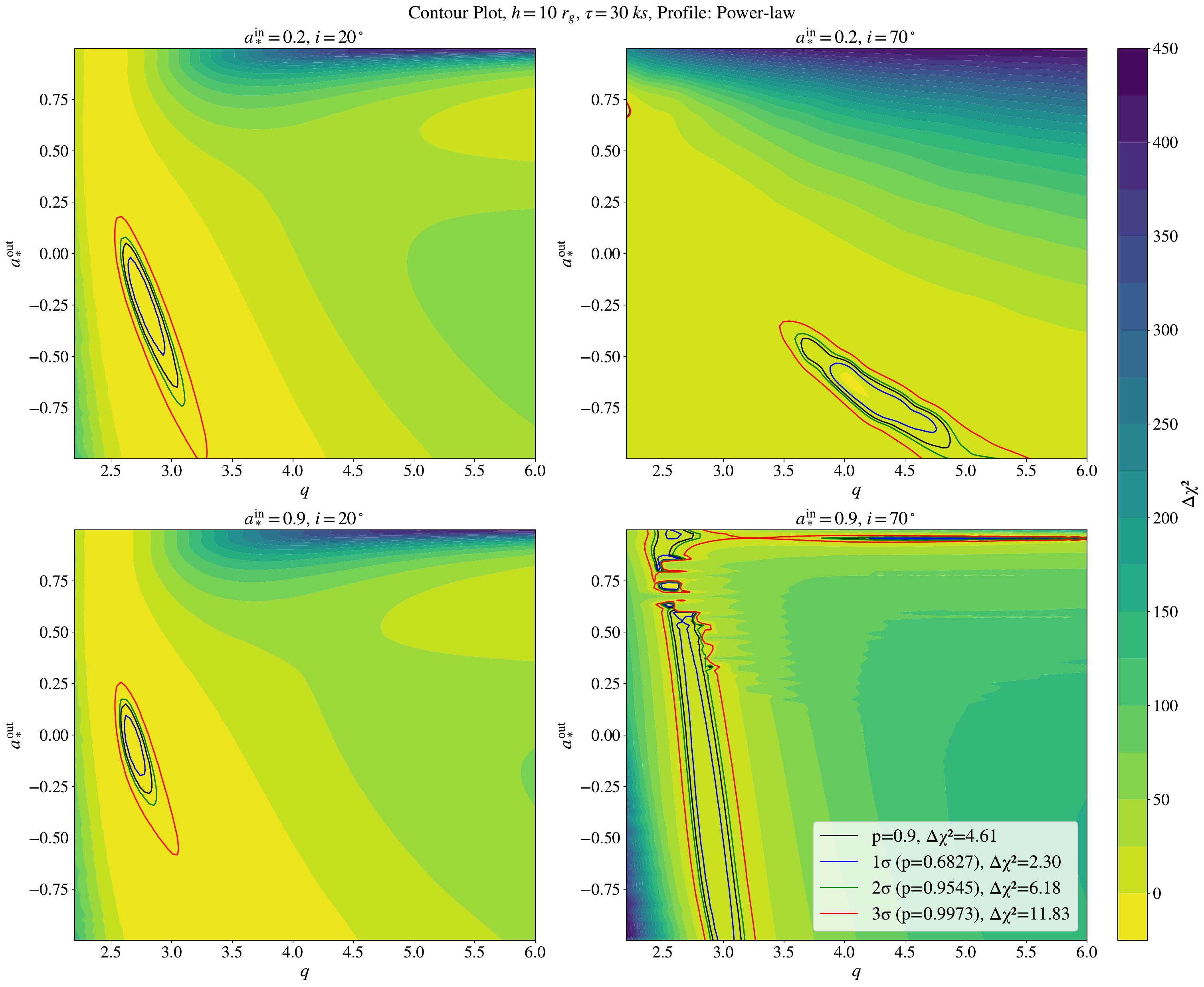}
	\vspace{-0.4cm}
	\caption{As in Fig.~\ref{contour_a_500} for some simulations with $h = 10~r_g$. The input spin is 0.2 (top panels) and 0.9 (bottom panels). The inclination angle of the disk is $20^\circ$ (left panels) and $70^\circ$ (right panels). The exposure time is $\tau = 30$ ks and the fits employ a simple power-law emissivity profile.}
	\label{contours}
\end{figure*}

When the corona is not close to the black hole (the case with $h = 10$~$r_g$), we may have even the opposite case, namely we may underestimate the black hole spin of some fast-rotating black holes. Fig.~\ref{contours} shows the contour plots of $\Delta\chi^2$ in the plane spin vs emissivity index for four cases with $h = 10$~$r_g$: as we can see from the two bottom panels, in the simulations with $a_*^{\rm in} = 0.9$ we can incorrectly infer low values of the spin parameter. This is consistent with the claim in Ref.~\cite{Dauser_2013MNRAS.430.1694D} concerning the analysis of broadened iron lines: a very broadened iron line is only possible when the value $h$ is low and that of $a_*$ is high, but a not very broadened iron line is possible either when the values of both parameters are high or when they are both low and it is difficult to distinguish the two cases.

\subsection{Disk-like Corona}

For the simulations for the disk-like corona with $R_{\rm disk,out} = 2$~$r_g$, similar to the findings reported in~\cite{relxilldisknk_2022ApJ...925...51R}, we recover the results of the lamppost simulations described in the previous subsection and therefore we do not show the plots here. Geometrically, the disk-like corona in this case is very compact and approaches the spatial scale characteristic of a point-like source.

\begin{figure*}[t]
	\centering
	\includegraphics[width=\linewidth]{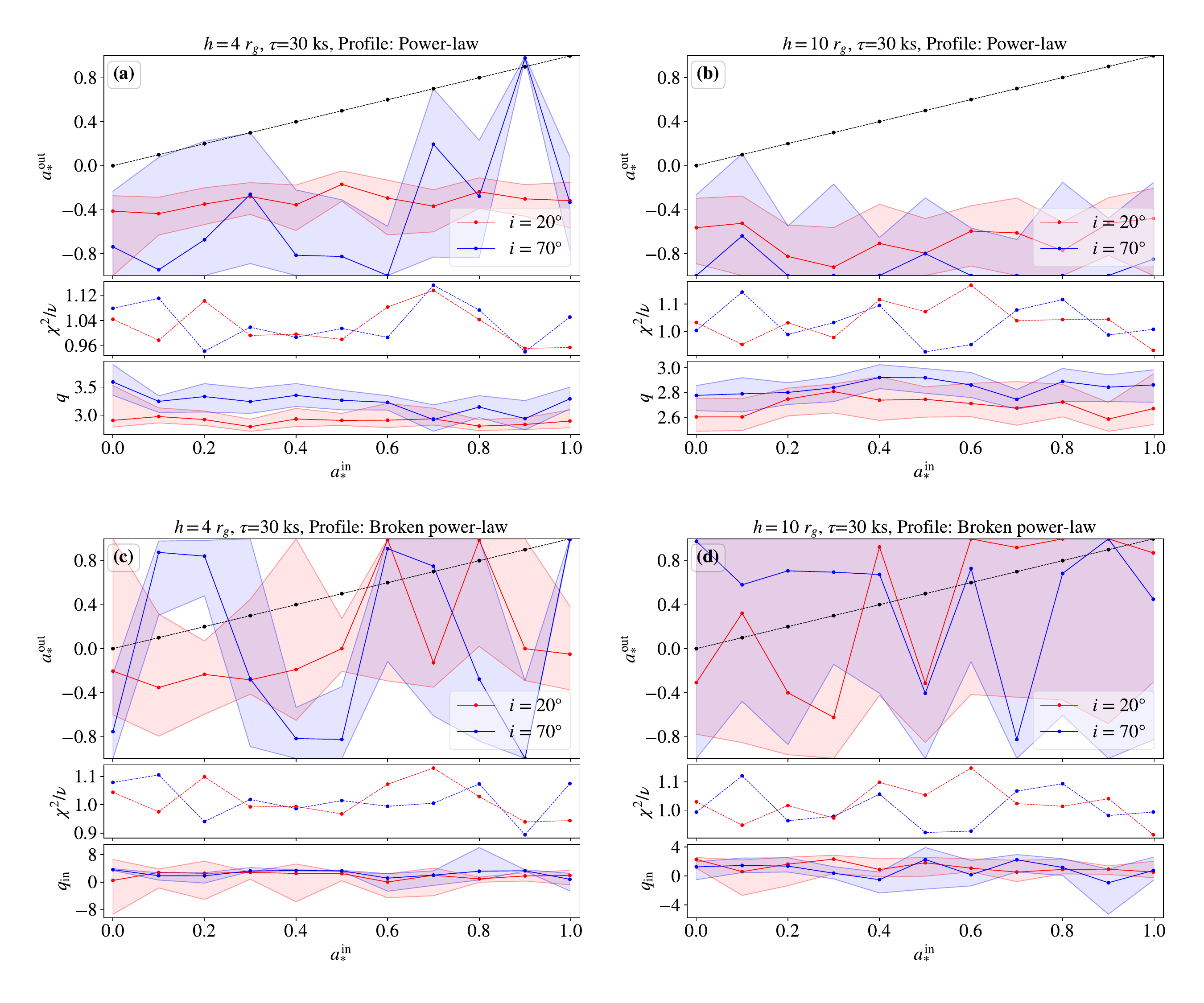}
	\vspace{-1.0cm}
	\caption{Results for the disk-like corona simulations with corona height $h=4$~$r_{g}$ (left panels) and $h=10$~$r_{g}$ (right panels) when the exposure time is $\tau=30$~ks. The corona outer radius is always $R_{\rm disk,out} = 20$~$r_g$. In the top panels, we fit the data with a power-law emissivity profile. In the bottom panels, we fit the data with a broken power-law emissivity profile and we impose $q_{\rm out} = 3$. In every panel, we show: input spin $a_*^{\rm in}$ vs output spin $a_*^{\rm out}$ (top quadrant), reduced $\chi^{2}$ of the best-fit model (central quadrant), and input spin vs best-fit emissivity index (bottom quadrant). The dots indicate the best-fit values and the shaded regions indicate the 90\% confidence intervals. Red is for the simulations with input inclination angle $i=20^\circ$ and blue is for the simulations with $i=70^\circ$. In the quadrants input spin vs output spin, we report in black the reference line $a_*^{\rm in} = a_*^{\rm out}$.}
	\label{nustar_30ks_disk}
\end{figure*}

The results of the simulations with $R_{\rm disk,out} = 20~r_g$ and exposure time 30~ks are summarize in Fig.~\ref{nustar_30ks_disk}. The input coronal height is $h = 4$~$r_g$ in the left panels and $h = 10$~$r_g$ in the right panels. We fit the simulated data with a simple power-law emissivity profile in the top panels and a broken power-law emissivity profile with $q_{\rm out} = 3$ in the bottom panels.

\begin{figure*}[t]
	\centering
	\includegraphics[width=\linewidth]{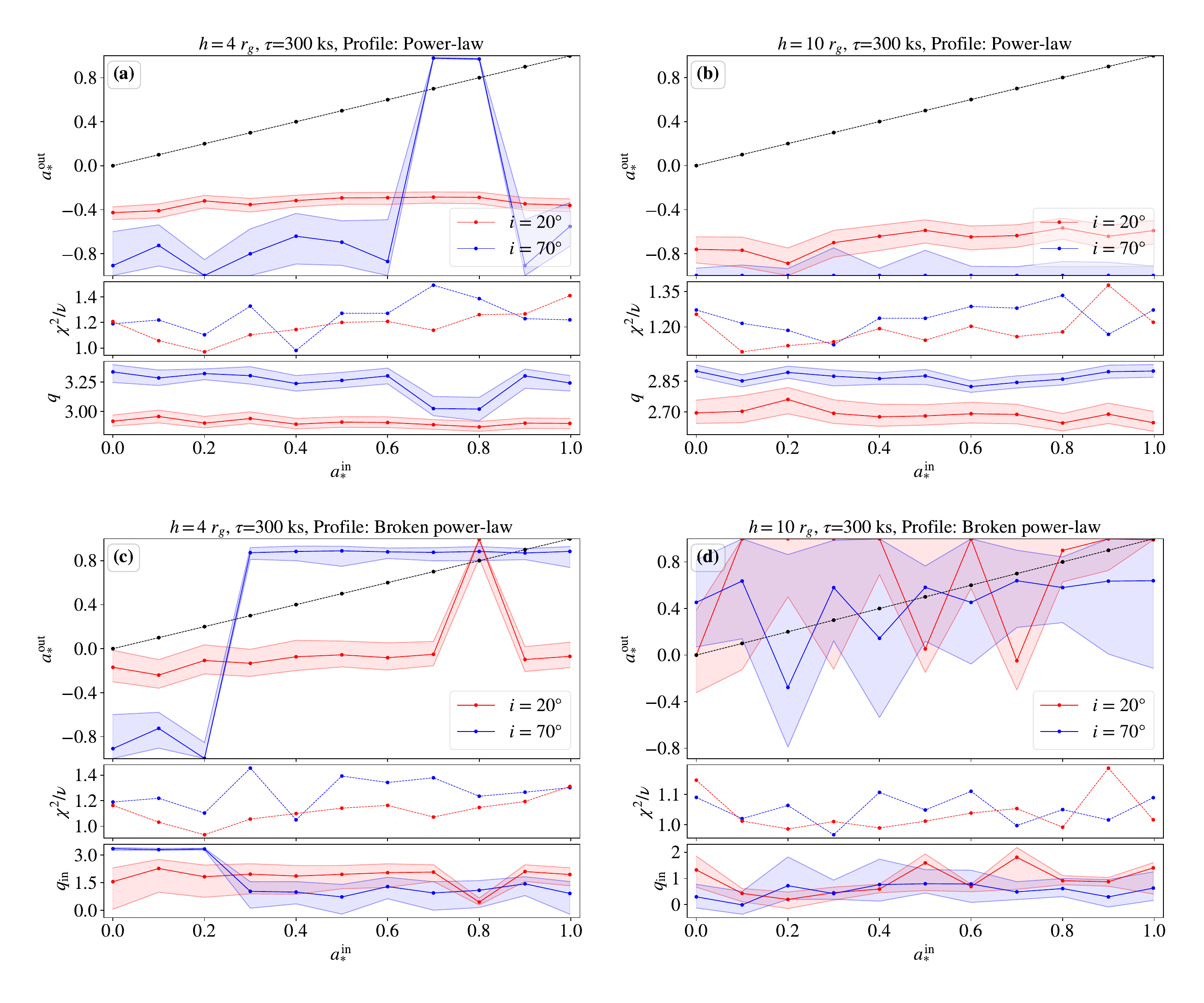}
	\vspace{-1.0cm}
	\caption{As in Fig.~\ref{nustar_30ks_disk} for simulations with exposure time $\tau=300$~ks.}
	\label{nustar_300ks_disk}
\end{figure*}

The results of the simulations with exposure time 300~ks are summarize in Fig.~\ref{nustar_300ks_disk}: as in Fig.~\ref{nustar_30ks_disk}, the input coronal height is $h = 4$~$r_g$ (left panels) and $h = 10$~$r_g$ (right panels) and the fits assume a simple power-law emissivity profile (top panels) and a broken power-law emissivity profile with $q_{\rm out} = 3$ (bottom panels).

For higher coronal heights ($h = 10~r_g$), spin measurements become even less accurate. In the broken power-law cases with fixed $q_{\rm out} = 3$, the confidence intervals widen so dramatically compared to the $h = 4~r_g$ cases that the model effectively fails to constrain spin. For the simple power-law cases, although the confidence intervals are not as extreme as in the broken power-law scenarios, the measured spin values deviate substantially from input values and often yield negative results. This systematic failure occurs across all input spin values, indicating that the model exhibits minimal sensitivity to the input spin value.

At lower coronal heights ($h = 4~r_g$), confidence intervals of spin are narrower than for the $h = 10~r_g$ cases, yet the problem of negative spin measurements persists. Both the simple power-law model and the broken power-law model with fixed $q_{\rm out} = 3$ fail to measure spin accurately, even with lower coronal heights and longer exposure times.

For both higher and lower coronal heights, the influence of inclination angles is not dominant but remains observable: higher inclinations reduce spin measurement accuracy and produce systematically more negative spin estimates. Regarding the influence of exposure time, longer exposure time ($\tau = 300$~ks) indeed narrow the confidence intervals of spin measurements. However, the increased exposure time does not reduce the systematic bias in these measurements.

In summary, these results may suggest that extended coronae require more sophisticated modeling, as neither simple nor broken power-law emissivity profiles can adequately describe their reflection features for reliable spin measurements. However, we note that our simulations only demonstrate that this is true for disk-like coronae, while other extended coronal geometries may still be reasonably approximated by phenomenological emissivity profiles.

\section{Discussion}\label{discussion}

In this section, we discuss our results. We would like to point out that our study is confined to examining how prescriptions for the emissivity profile affect spin measurements. In practice, additional assumptions and simplifications within reflection models can also influence the robustness of these measurements~\cite{Bambi_2021SSRv..217...65B}. Such modeling effects are expected to become the limiting factor for the next generation of X-ray observatories (e.g., eXTP~\cite{2025SCPMA..6819502Z}), which promise to provide unprecedented high-quality data. This motivates both a more accurate assessment of the impact of these assumptions and the development of more advanced reflection models.

\subsection{Iron Line Shape}\label{iron line shape}

\begin{figure}
	\centering
	\includegraphics[width=1.\linewidth]{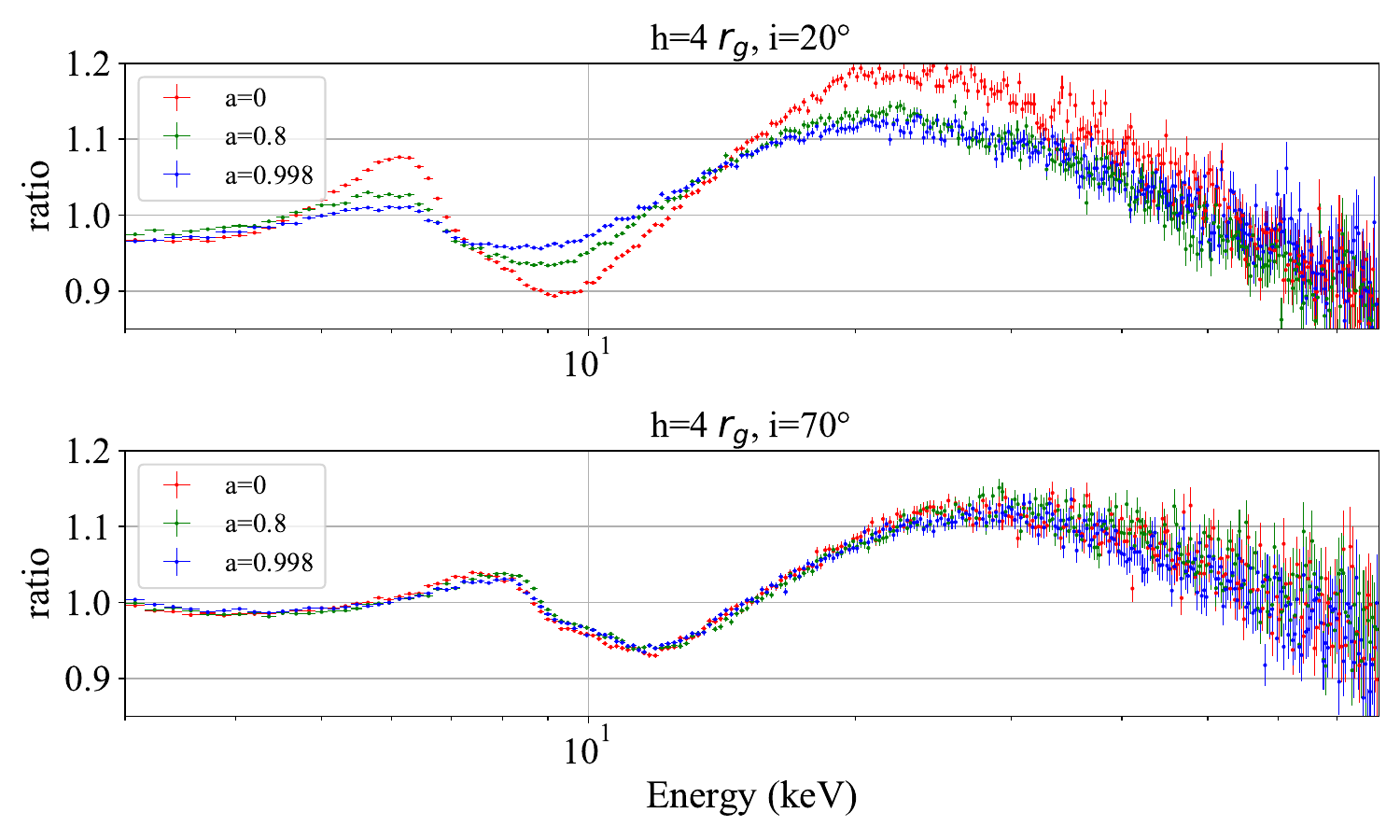}
	\caption{Data-to-model ratio plots for the lamppost corona geometry data with $h = 4$~$r_{g}$ and exposure time $\tau = 30$~ks, highlighting residuals relative to the best-fit of power-law continuum model in XSPEC. Upper panel: Inclination $i = 20^\circ$. Lower panel: Inclination $i = 70^\circ$. Color-coded spin values: Red: Input spin $a_*^{\text{in}} = 0$; Green: Input spin $a_*^{\text{in}} = 0.8$; Blue: Input spin $a_*^{\text{in}} = 0.998$.}
	\label{iron_line_30ks_h_4_lamppost}
\vspace{0.5cm}
	\centering
	\includegraphics[width=1.\linewidth]{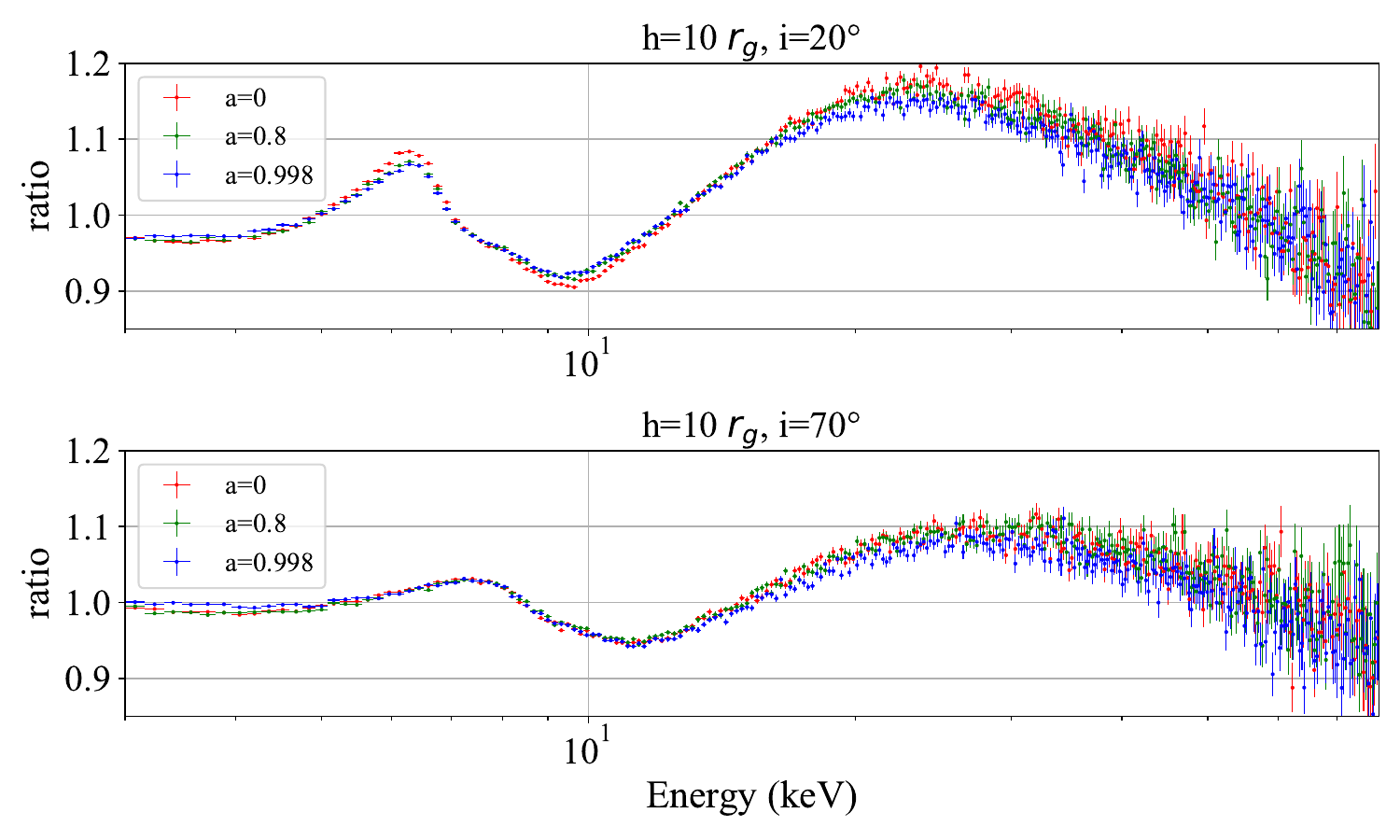}
	\caption{Data-to-model ratio plots for the lamppost corona geometry data with $h = 10$~$r_{g}$ and exposure time $\tau = 30$~ks, highlighting residuals relative to the best-fit of power-law continuum model in XSPEC. The coordinate axes, line styles, and color schemes in this figure adhere to identical conventions as defined in Fig.~\ref{iron_line_30ks_h_4_lamppost}}
	\label{iron_line_30ks_h_10_lamppost}
\end{figure}

In spin measurements, the broadness  of the iron line plays a crucial role. To understand why spin constraints vary across cases, we compare the iron line broadening. By fitting the spectra with a power-law continuum model in XSPEC, we manifest iron line features. Fig.~\ref{iron_line_30ks_h_4_lamppost} demonstrates that at low coronal heights, iron line broadness exhibits greater sensitivity to spin at low inclination angles than at high inclinations. Indeed, if the inclination angle is low, the effect of Doppler boosting is negligible, so the broadening of the iron line is determined by the gravitational redshift, which, in turn, depends only on the black hole spin. If the inclination angle is high, he broadening of the iron line is determined by the combination of the Doppler boosting and gravitational redshift, and it is difficult to disentangle these two effects without an {\it a priori} knowledge of the emissivity profile of the disk. This explains the larger spin uncertainties at high inclinations, as Doppler boosting (enhanced at higher inclination angle) overwhelms gravitational redshift (enhanced at higher spin) to a greater extent at higher inclination angles. The differences in line broadening induced by spin are more pronounced at smaller inclination angles. At high inclinations, even the iron line profile with an $r^{-3}$ emissivity for negative spin values shows limited distinction from that of high positive spin (see Ref.~\cite{Dauser_2010MNRAS.409.1534D}).

For higher coronal heights, Fig.~\ref{iron_line_30ks_h_10_lamppost} demonstrates decreased sensitivity of line broadness to spin compared to lower heights, as photons experience stronger gravitational effects when the corona is closer to the black hole. These accounts for the larger uncertainties in spin measurements for cases with high inclination angles in section~\ref{results_lamppost}.
Indeed, higher coronal positions reduce light bending, weakening illumination of the inner disk, and then diminishing relativistic effects like iron line broadening and redshift. Consequently, spin-dependent spectral features become less distinct.
Some studies suggest that the position of the corona significantly affects the illumination pattern of the accretion disk~\cite{relxilldisknk_2022ApJ...925...51R}, returning radiation effect~\cite{Dauser_2022MNRAS.514.3965D, Riaz_2023EPJC...83..838R}, and the observed flux~\cite{Feng_2025ApJ...984..173F}.

\begin{figure}
	\centering
	\includegraphics[width=1.\linewidth]{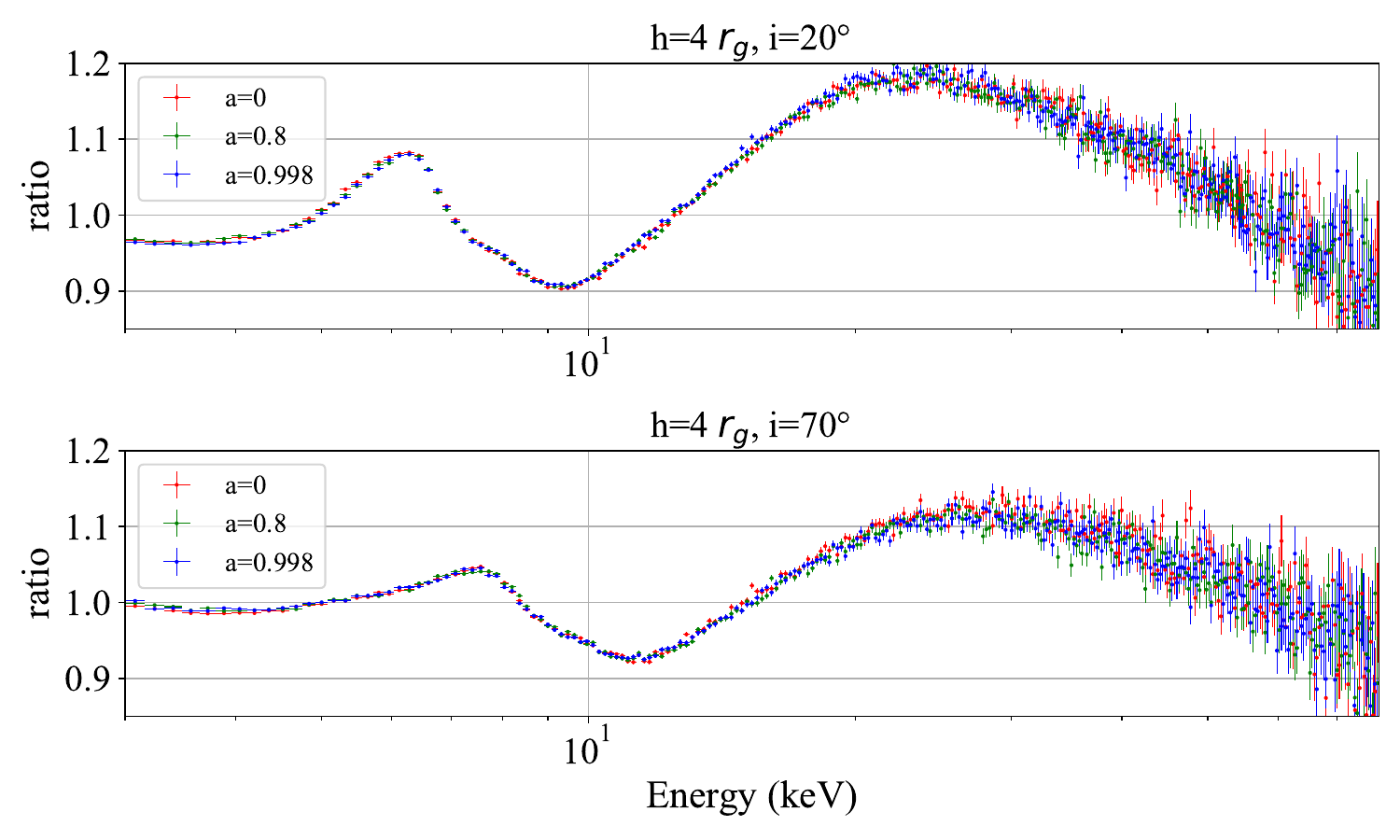}
	\caption{Data-to-model ratio plots for the disk-like corona geometry with $h = 4$~$r_{g}$ and exposure time $\tau = 30$~ks, highlighting residuals relative to the best-fit of power-law continuum model in XSPEC. The coordinate axes, line styles, and color schemes in this figure adhere to identical conventions as defined in Fig.~\ref{iron_line_30ks_h_4_lamppost}}
	\label{iron_line_30ks_h_4_disk}
\end{figure}

Regarding the disk-like corona scenario, Fig.~\ref{iron_line_30ks_h_4_disk} reveals negligible differences in iron line broadness between low and high inclinations even at low coronal heights, then we are not presenting iron-line plots for higher coronal height in this case. In the extended disk-like corona (with $R_{\rm disk,out} = 20$~$r_g$) scenario, the iron line feature of the spectrum is not sensitive to the spin. 
This can be explained by the fact that as the corona's outer rim expands away from the black hole, it more effectively illuminates the disk at larger radii. Consequently, the relative contribution of emission from the innermost region -- which is the most sensitive to spin -- to the total spectrum decreases~\cite{relxilldisknk_2022ApJ...925...51R}.

In summary, spin measurements exhibit greater uncertainty for observations at high inclination angles. This conclusion is supported by Ref.~\cite{Du_2024ApJ...963..152D}, where the authors also find that larger inclinations lead to less constrained spins, and by Ref.~\cite{Cardenas_2020PhRvD.101l3014C}, where the authors note that systematic uncertainties and modeling biases are more severe for low-spin sources observed at high inclination.

Lastly, we note that future calorimeter missions (e.g., NewAthena/X-IFU) will become increasingly sensitive to subtle deviations in the iron line profile. This can further challenge the capability of phenomenological emissivity profiles to recover correct measurements.


\subsection{Power-law Emissivity Profile}

In reflection spectra, the emissivity profile encodes the corona's illumination pattern on the accretion disk and the effects of the gravitational field on photon trajectories~\cite{Miniutti_2004MNRAS.349.1435M, Wilkins_and_Fabian_2012MNRAS.424.1284W}. Since the spin determines the location of the ISCO, a higher spin moves the inner edge of the accretion disk closer to the strong-gravity region~\cite{Dokuchaev_2019Univ....5..183D}, resulting in a steeper emissivity profile~\cite{Svoboda_2012A&A...545A.106S}. For a truncated inner disk with a spin approaching $a_* \sim 0.998$, it is possible to fit the spectrum using a relatively flat emissivity index ($q \sim 3$)~\cite{Gallo_2022MNRAS.515.2208G}.  
Observational results frequently reveal steep emissivity profiles in the inner disk regions (e.g., Refs.~\cite{Xu_2018ApJ...865..134X, Zhang_2019ApJ...884..147Z, Dong_2020MNRAS.493.2178D, Tripathi_2021ApJ...913...79T, Zhang_2021PhRvD.103b4055Z, Tripathi_2021ApJ...907...31T}). A compact lamppost-like corona located near the black hole ($h \sim$~a few~$r_g$) can strongly illuminate the innermost disk and produce such steep profiles~\cite{Svoboda_2012A&A...545A.106S}. Furthermore, ionization gradients and electron density gradients in the disk may also affect the emissivity profile~\cite{Shreeram_2020MNRAS.492..405S, Abdikamalov_2021PhRvD.103j3023A, Mall_2022MNRAS.517.5721M}. The presence of a radial ionization gradient could further increase the measured emissivity index~\cite{Abdikamalov_2021PhRvD.103j3023A}.

\begin{figure*}
	\centering
	\includegraphics[width=1\linewidth]{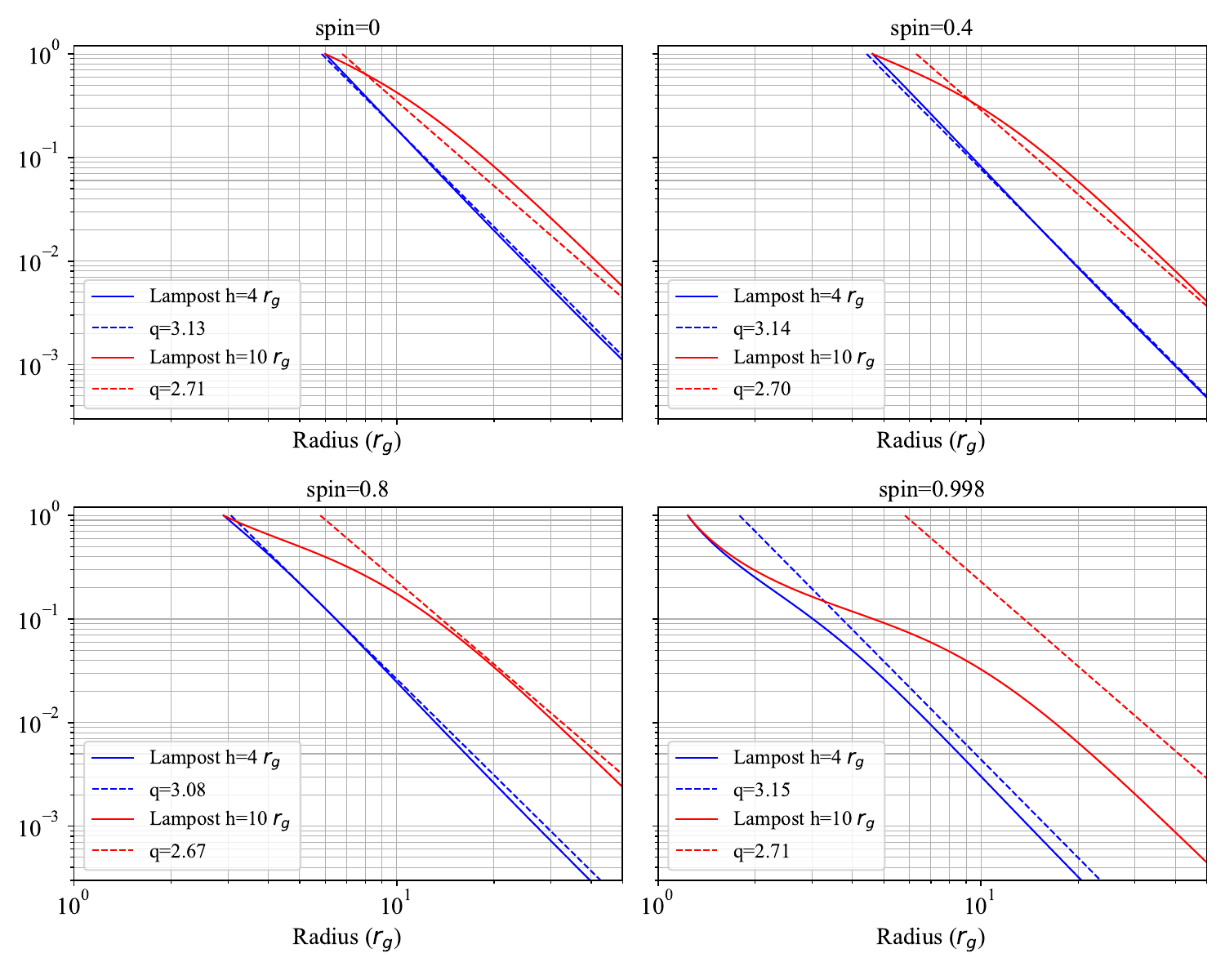}
	\vspace{-0.7cm}
	\caption{Comparison of emissivity profiles for lamppost coronae at different spin values and coronal heights. Blue and red solid lines show the theoretical lamppost emissivity profiles for $h=4~r_g$ and $h=10~r_g$ coronae, respectively. Blue and red dashed lines represent the best-fit power-law emissivity profiles obtained from fits to simulated spectra with $i=20^\circ$ and exposure time $\tau = 300$ ks for $h=4~r_g$ and $h=10~r_g$ coronae, respectively.}
	\label{lamppost_emissivity_compare}
\end{figure*}

In the fitting results of this paper, for a lamppost corona at lower heights ($h = 4~r_g$), we find that fitting with a power-law emissivity profile (fewer free parameters) yields more accurate spin estimates than using a broken power-law model (compare the top left panel with the bottom left panel in Fig.~\ref{nustar_30ks_lampost}). An explanation for this may be that the emissivity profile of a lower heights lamppost corona closely follows a simple power-law (Fig.~\ref{lamppost_emissivity_compare}, blue lines for $h = 4~r_g$) without significant breaks within $R < 20~r_g$. 
When fitting with a broken power-law, the extra parameter $q_{\rm out}$ increases uncertainties in both $q_{\rm in}$ and $q_{\rm out}$ without improving precision. In spectral data analysis, the introduction of additional parameters may lead to parameter degeneracy. In our study, fixing $q_{\rm out} = 3$ restores accuracy comparable to the simple power-law case in higher quality of data (compare the central left panel with the bottom left panel in Fig.~\ref{nustar_300ks_lampost}).

At higher coronal heights ($h = 10~r_g$), fixing $q_{\rm out} = 3$ yields no significant improvement. While the broken power-law model formally outperforms the simple power-law for spin fitting, its constraints remain poor: confidence intervals for spin are excessively wide, preventing accurate measurements. 

\begin{figure*}
	\centering
	\includegraphics[width=1\linewidth]{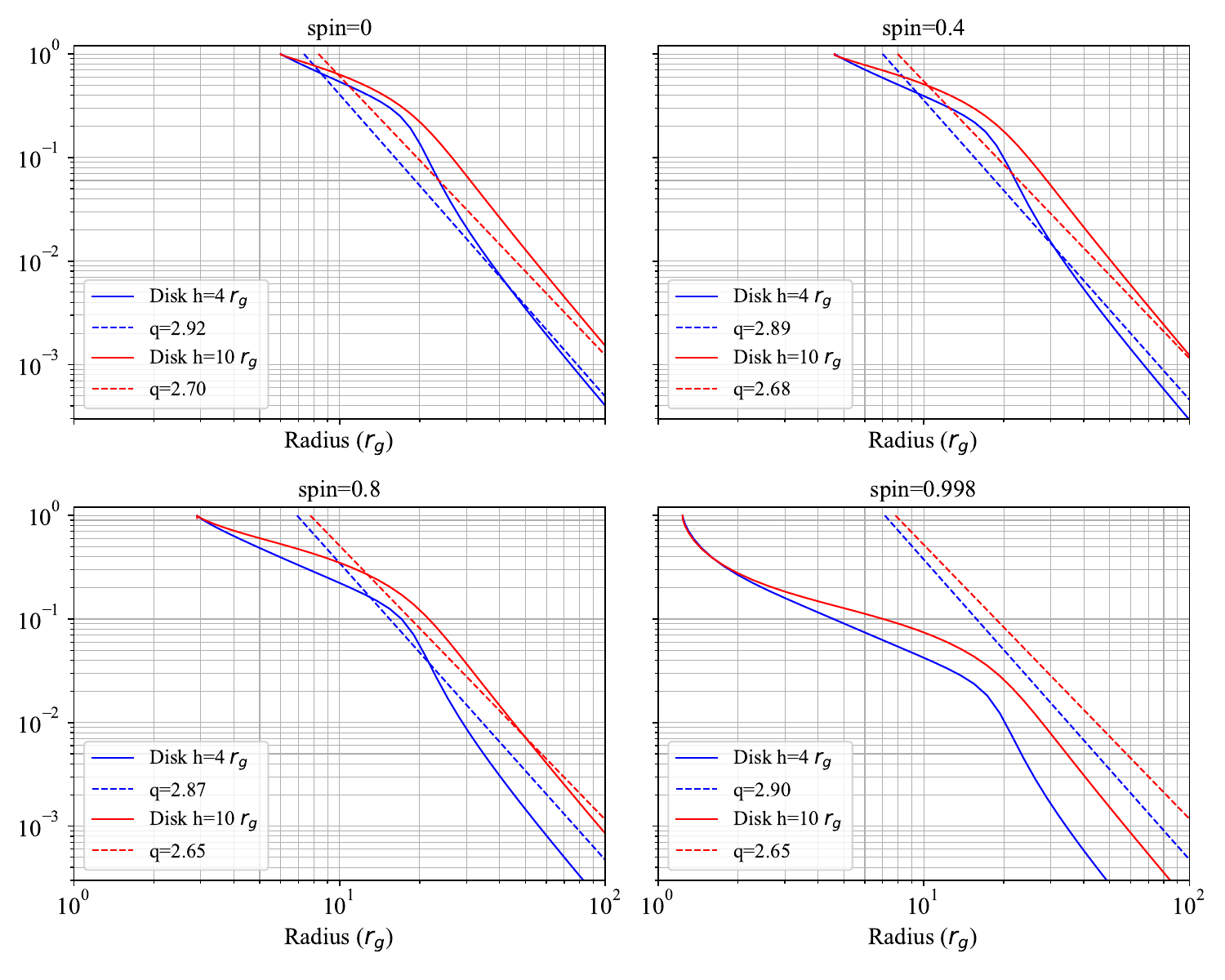}
	\vspace{-0.7cm}
	\caption{Comparison of emissivity profiles for disk-like coronae with $R_{\rm disk,out} = 20$ at different spin values and coronal heights. Blue and red solid lines show the theoretical disk-like emissivity profiles for $h=4~r_g$ and $h=10~r_g$ coronae, respectively. Blue and red dashed lines represent the best-fit power-law emissivity profiles obtained from fits to simulated spectra with $i=20^\circ$ and exposure time $\tau = 300$ ks for $h=4~r_g$ and $h=10~r_g$ coronae, respectively.}
	\label{disk_emissivity_compare}
\end{figure*}

Fig.~\ref{lamppost_emissivity_compare} confirms that at $h=10~r_g$, the emissivity profile deviates significantly from both simple and broken power-laws, indicating the models' poorer performance relative to $h=4~r_g$. For disk-like corona, spin measurements using power-law emissivity profiles remain unsatisfactory even for lower coronal height. Fig.~\ref{disk_emissivity_compare} shows that the emissivity profile of disk-like coronae deviates significantly from a power-law shape; this also indicates that the models exhibit poorer performance in disk-like corona cases.  

In summary, when the corona is close to the black hole both vertically and radially, we recommend starting with a simple power-law emissivity profile to avoid parameter degeneracies inherent in broken power-law models. For distant corona, spin measurements using power-law or broken power-law emissivity profiles yield unsatisfactory results.
It should be further noted that although the simple or broken power-law emissivity profile does not perform well in some of the scenarios above, neither the lamppost model nor the disk-like corona may always accurately reflect the true geometric configuration of the corona in real observations. In Ref.~\cite{Jiang_2022MNRAS.514.3246J}, the authors suggest that the lamppost corona geometry tends to overestimate the reflection fraction parameter, particularly when accounting for the thickness of the accretion disk. The study in Ref.~\cite{Shashank_2025arXiv250702583S} based on GRMHD simulations points out that, if the corona is modeled as the base of the jet, the broken power-law emissivity profile also performs better than the lamppost emissivity profile.

\subsection{Exposure Time and data quality}

In X-ray spectroscopy analysis, extending the exposure time directly enhances the statistical signal-to-noise ratio (S/N) of the spectrum. According to Poisson statistics (the default method used by XSPEC for generating simulated data~\cite{xspec_manual}), the signal-to-noise ratio scales with exposure time $\tau$ as S/N $\propto\sqrt{\tau}$. We find that the improved S/N from longer exposures effectively reduces the statistical uncertainties of the spin $a_{*}$ and emissivity index $q$, the width of the confidence interval for $a_{*}$ and $q$ decreases significantly (e.g., compare the top left panel in Fig.~\ref{nustar_30ks_lampost} with the top left panel in Fig.~\ref{nustar_300ks_lampost}). However, in cases where measurements exhibit substantial systematic deviations from input values at 30~ks exposure, increasing the exposure time to 300~ks does not significantly mitigate this bias (e.g., compare the top right panel in Fig.~\ref{nustar_30ks_lampost} with the top right panel in Fig.~\ref{nustar_300ks_lampost}).

In X-ray spectroscopy, improved data quality enables more precise measurements of black hole spin parameters~\cite{Reynolds_2021ARA&A..59..117R, Bambi_2021SSRv..217...65B}. 
With high-quality data, the quality of the fit can help determine whether a power-law emissivity profile provides an adequate description of the reflection pattern~\cite{Zhang_2019ApJ...884..147Z, Liu_2019PhRvD..99l3007L}.
The next generation of X-ray detectors with their significantly enhanced data quality, will expect more sophisticated modeling for reflection spectra~\cite{Dong_2023arXiv231209210D, 2025MNRAS.536.2594L, Huang_2025PhRvD.111f3025H, Huang_2025ApJ...989..168H}.

\subsection{Impact of the energy resolution of the instrument}

\begin{figure*}[t]
	\centering
	\includegraphics[width=\linewidth]{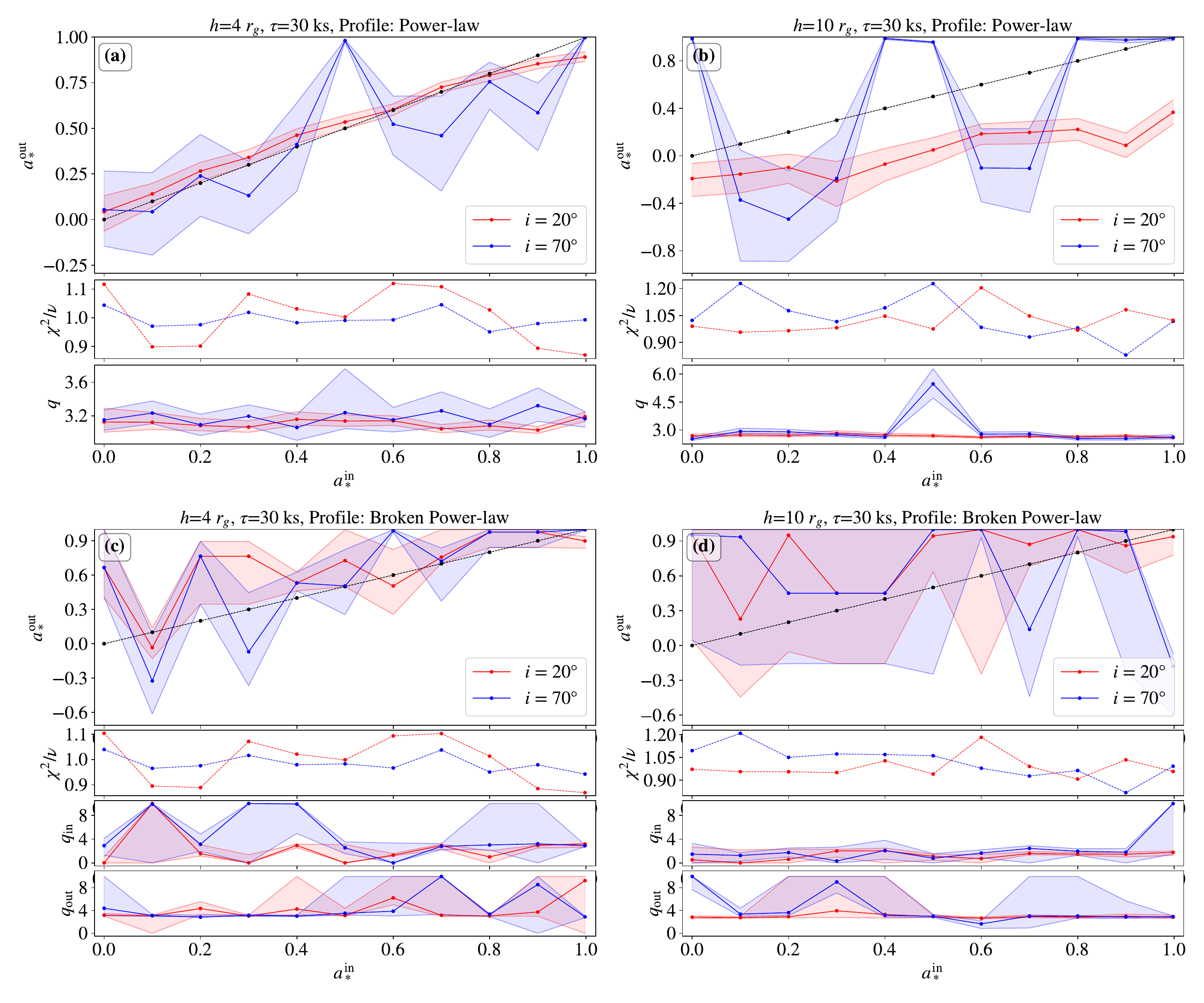}
	\vspace{-0.8cm}
	\caption{As in Fig.~\ref{nustar_30ks_lampost} for simulations NICER+NuSTAR with exposure time of 5~ks for NICER and 30~ks for NuSTAR. We fit the simulated data with a model employing a simple power-law emissivity profile (top panels) and a broken power-law emissivity profile (bottom panels).}
	\label{nustar_nicer}
\end{figure*}

In our study, we have only considered simulations of NuSTAR observations, which have the advantage of detecting both the iron line in the soft X-ray band and the Compton hump in the hard X-ray band but have relatively low spectral resolution. In this subsection, in order to assess the impact of the energy resolution of the instrument, we simulate some observations NICER+NuSTAR, as NICER has a good energy resolution at the iron line. We repeat the simulations with a lamppost corona with $h=4~r_g$ and $h=10~r_g$ assuming an exposure time of 5~ks for NICER and 30~ks for NuSTAR. We fit the data with a simple power-law emissivity profile and a broken power-law emissivity profile. The results of the spin measurements are reported in Fig.~\ref{nustar_nicer} and should be compared with the results in Fig.~\ref{nustar_30ks_lampost}. While we see some improvement in the spin measurements for $h=4~r_g$, in the simulations with $h=10~r_g$ we only reduce the statistical uncertainties without improving the accuracy of the estimate of the black hole spins. Thus, a higher spectral resolution cannot compensate for the systematic uncertainties inherent in the astrophysical model.

\section{Conclusions}
\label{conclusion}

When coronal geometry is unknown, the power-law emissivity profile offers an efficient but potentially unreliable fitting approach. For compact, low-height corona and low-inclination systems, this profile can accurately measure spin using long-exposure data. Conversely, for high-height or high-inclination systems, spin measurements derived from the power-law emissivity profile lack both accuracy and reliability. Successful application requires the corona to be compact and close to the black hole. 

When employing a broken power-law, one must exercise caution by comparing results with those from the simple power-law model. For lamppost-dominated sources, if it is also close to black hole, the single power-law emissivity profile -- though physically simplistic -- yields more robust spin estimates than broken power-law models by avoiding artificial parameter degeneracies. The breaking radius $R_{\rm br}$ should be introduced when justified by independent physical evidence.

Extending exposure time improves measurement precision by reducing statistical errors in parameter estimation. However, it does not substantially enhance measurement validity, as it fails to mitigate systematic biases inherent to the model. 

In conclusion, we can translate the results of our simulations into practical advice for observers as follows. Black hole spin measurements inferred with phenomenological emissivity profiles are reliable only when the measurement of the spin is {\it high} and {\it precise} and, at the same time, the inferred value of the disk inclination angle is {\it low}. When one of these three ingredients is missing, it is difficult to distinguish accurate spin measurements from inaccurate ones. For example, if the measurement of the spin is low and precise and we infer a low value of the inclination angle, we cannot exclude that the source is a fast-rotating black hole but the corona does not illuminate well the inner part of the disk (the corona is not close to the black hole and/or is not compact). If the measurement of the spin is high but not precise, the source may be a slow-rotating black hole, regardless of the inferred value of the inclination angle of the disk. If the measurement of the spin is high and precise, but we also measure a high value of the inclination angle of the disk, we cannot exclude that the source is a slow-rotating black hole with a compact corona close to the black hole.

Last, we stress that our study is focused on the ability of phenomenological emissivity profiles to model the actual illumination of the disk by coronae of specific geometry and we have ignored other inevitable complications in the analysis of real observations. Our simulations are based on an absorbed spectrum with a Comptonized continuum and a relativistically broadened reflection spectrum. In reality, the spectrum may have also a non-relativistic reflection component that can further challenge the ability of a reflection model to recover the correct values of the parameters of a system.


\vspace{0.5cm}

{\bf Acknowledgments --}
This work was supported by the National Natural Science Foundation of China (NSFC), Grant No.~W2531002.


\bibliography{bibliography} 

\end{document}